\documentstyle[twocolumn,aps,prl,epsfig]{revtex}

\begin{document} 
  \draft
\twocolumn[\hsize\textwidth\columnwidth\hsize\csname
@twocolumnfalse\endcsname

\title{Interlayer Coupling and p-wave Pairing in Strontium Ruthenate}

\author{James F. Annett$^1$, G. Litak$^2$, B. L. Gy\"orffy$^1$, K. I. Wysoki\'nski$^3$\\ 
$^1$H. H. Wills Physics Laboratory, University of Bristol, Tyndall Ave, 
BS8-1TL, UK.\\ 
$^2$ Department of Mechanics, Technical University of Lublin,\\ 
Nadbystrzycka 36, 20-618 Lublin, Poland.\\ 
$^3$ Institute of Physics, M. Curie-Sk\l odowska University,\\ 
Radziszewskiego 10, 20-031 Lublin, Poland} 
\date{\today} 
\maketitle  
 
\begin{abstract} 
On the basis of a three orbital model and an effective attractive 
interaction between electrons  we investigate the 
possible superconducting states, with $p$ and $f$-wave internal symmetry, of 
Sr$_2$RuO$_4$.
For an orbital dependent interaction which acts between in plane and out of
plane nearest neighbour Ruthenium atoms we find a state for which the gap in
the quasi-particle spectra has a line node on the $\alpha $ and 
$\beta $ sheets of
the Fermi Surface, but it is complex
with no nodes on the $\gamma$-sheet. We show that this
state is consistent with all the available  experimental data. In
particular, we present the results of our calculations of the specific
heat and penetration depth as functions of the temperature.
\end{abstract} 

\pacs{PACS numbers:
             74.70.Pq,    
             74.20.Rp,    
             74.25.Bt     
}

\vspace{5mm}
]
\narrowtext


Ever since triplet pairing was discovered in superfluid $^{3}$He\cite{legg},
during the
early seventies, there has been a constant search for the superconducting
analogue of this intriguing macroscopic quantum phenomenon.
Although, as
yet, there is no metal for which triplet paring has been definitively
demonstrated there are a number of good candidates. 
The evidence that Sr$_{2}$RuO$_{4}$ is a triplet superconductor is 
particularly strong\cite{maeno,mac}. 
 Nevertheless, even in this case the full symmetry of the equilibrium state
below T$_{c}$ remains open to 
debate\cite{maeno,mac,agte,maz,miya,graf,won,dahm,sig,hase,zhit}.

One of the puzzles currently at the centre of attention is the apparent
incompatibility between experimental evidence for broken time-reversal
symmetry in the superconducting state\cite{luke,sig}
 and equally convincing measurements
indicating that the order parameter ${\bf d}({\bf k})$ has a
 line of nodes on the Fermi
Surface\cite{nishi,DalevH}.  
The reason why this state of affairs represents a dilemma is
that for all odd parity spin triplet pairing states in tetragonal crystals,
 group theory does not require the simultaneous presence of both broken
time-reversal symmetry and line nodes\cite{annett}. 
Consequently, due to their lower
condensation energy, line nodes are unlikely.  For instance, the pairing
state 
${\bf d}({\bf k}) \sim (k_{x}+ik_{y})\hat{\bf e}_{z}$, proposed by 
Agterberg {\it et al.}\cite{agte} on the grounds that it
minimizes the free energy, obviously breaks time reversal invariance and has
no line nodes.  Of course, such states as
${\bf d}({\bf k})\sim (k_{x}+ik_{y})(k_{x}k_{y})\hat{\bf e}_{z}$
discussed by Graf and Balatsky\cite{graf}
and other f-wave states\cite{won,dahm,hase} are
allowed by symmetry considerations but, the point is that, the nodes
are not required by symmetry and hence it is  not very attractive 
ansatz to build a theory on.
Under these circumstances it is more advantageous to study physically motivated
microscopic models even if the question of the actual mechanism of pairing
is to be avoided. In this letter we propose and investigate one such
physically motivated model.

Our model is prompted by the observation, of Hasegawa {\it et al.}\cite{hase}, 
that coupling between the Ruthenium layers 
in Sr$_{2}$RuO$_{4}$ leads to convenient,
horizontal, $k_{z}=\pm \frac{\pi }{c}$ lines of
zeros on the Fermi Surface.  It features two,  intra and  inter plane,
interaction constants $U_{\parallel }$ and $U_{\perp }$
respectively, which describe the attraction between electrons each occupying
one of three, t$_{2g},$ orbitals on  Ru atoms which are nearest neighbours 
either
in plane and out of plane. As we shall show, this simple physical
picture yields horizontal nodes on the $\alpha $,$\beta $ sheets of the
Fermi Surface whilst the $\gamma $-sheet is fully gaped with 
${\bf d}({\bf k}) \sim (\sin{k_{x}}+i\sin{k_y})\hat{\bf e}_z$, in 
quantitative as well as qualitative agreement with
a number of experiments.

Hasegawa {\it et al.}\cite{hase}, treated the case of 
 a single band only and, unlike us, they
made no quantitative contact with experiments.
The more recent work of Zhitomirsky and Rice\cite{zhit} is closer to
ours, although dealing with a two
band model only. They also couple electrons on different layers and 
find line nodes on
the $\alpha ,\beta $ sheets but a gap on $\gamma$. However, the physics
behind their model, as well as its consequences, are  quite different from
ours.  Largely, this is due to a difference in strategy.  In a multi-band BCS
like model, with different coupling constants for each band, one generically
finds multiple phase transitions as the different sheets of the Fermi
Surface are gaped on lowering the  temperature. Since experimentally there is
only one jump in the specific heat, at T$_{c}=1.5K$, in constructing a
sensible model one must tailor it to eliminate such double
transitions. Zhitomirsky and Rice\cite{zhit} chose to couple the 
order parameters on 
different sheets of the Fermi Surface.  This hybridized the different order
parameters and led to a single transition. They referred to such hybridization
as a inter-band proximity effect.   We, on the other hand, adjusted the two
coupling constants, $U_\parallel $ and $U_\perp $, so that the transition on
the $\alpha ,\beta $ sheets would occur at  more or less the same
temperature as as on $\gamma $. Somewhat surprisingly these two approaches
imply different physics. The proximity model requires a three point interaction
to mix the in-plane and out of plane Cooper pairs whilst our ``local  bond''
model is a strictly a two point effect. Moreover, the ``local bond'' model has
fewer free parameters by construction and therefore, as will be shown
below, is more readily compared with experiments.

To describe the superconducting state we employ a simple multi-band
attractive Hubbard model: 
\begin{eqnarray}
  \hat{H}& =& \sum_{ijmm',\sigma}  
\left( (\varepsilon_m  - \mu)\delta_{ij}\delta_{mm'}  
 - t_{mm'}(ij) \right) \hat{c}^+_{im\sigma}\hat{c}_{jm'\sigma} \nonumber \\
&& - \frac{1}{2} \sum_{ijmm'\sigma\sigma'} U_{mm'}^{\sigma\sigma'}(ij)
 \hat{n}_{im\sigma}\hat{n}_{jm'\sigma'} \label{hubbard}
\end{eqnarray}
 where $m$ and 
$m'$ refer to the three Ruthenium $t_{2g}$ orbitals $a=xz$, 
$b = yz$ and $c = xy$ and  $i$ and $j$ label the sites  
of a body centered tetragonal lattice.
The hopping integrals $t_{mm'}(ij)$ and site 
energies $\varepsilon_m$ were fitted to reproduce the experimentally 
determined Fermi Surface \cite{berg}.
The set of interaction 
constants $U_{mm'}^{\sigma\sigma'}(ij)$ describe attraction 
between electrons on nearest neighbour sites
with spins $\sigma$ and $\sigma'$
and in orbitals $m$ and $m'$.
Thus our actual calculations consists of solving, self-consistently, 
the following Bogoliubov-de Gennes equation: 
\begin{equation} 
 \sum_{jm'\sigma'} \left(\begin{array}{c} 
 E^\nu - H_{mm'}(ij)  ~ ~  ~
 \Delta^{\sigma\sigma'}_{m,m'} (ij)\\ 
 \Delta^{\sigma\sigma'}_{mm'}(ij) ~ ~ ~ ~
 E^\nu +  H_{mm'}(ij)
\end{array}\right) 
\left(\begin{array}{ll} 
 u^\nu_{j m'\sigma'}\\ 
v^\nu_{jm'\sigma'}\end{array}\right)=0\,, \label{bogoliubov}
\end{equation} 
where  $ H_{mm'}(ij) $ 
is the normal spin independent part of the Hamiltonian, and  
the $\Delta^{\sigma\sigma'}_{mm'}(ij)$ is  
self consistently given 
in terms of the pairing amplitude, or order parameter, 
$\chi_{mm'}^{\sigma\sigma'}(ij)$, 
\begin{equation} 
 \Delta^{\sigma\sigma'}_{mm'}(ij) = U_{mm'}^{\sigma\sigma'}(ij) 
\chi_{mm'}^{\sigma\sigma'}(ij)\,. \label{deltas}
\end{equation} 
defined by the usual relation 
\begin{equation} 
\chi_{mm'}^{\sigma\sigma'}(ij) = 
\sum_{\nu} u^\nu_{im\sigma}v^\nu_{jm'\sigma'}
(1 - 2f(E^\nu))\,, 
\end{equation} 
where $\nu$ enumerates the solutions of Eq.~\ref{bogoliubov}.


We solved the above system of Bogoliubov de Gennes equations
including all three bands and the 
experimental three dimensional Fermi surface.
We assumed that the pairing interaction $U_{mm'}^{\sigma\sigma'}(ij)$
for nearest neighbours in plane is only acting 
for the $c$  ($d_{xy}$) Ru orbitals. We further
assumed that the nearest neighbour inter plane interaction
acts only in $a$ and $b$ orbitals  ($d_{xz}$, $d_{yz}$).
The motivation for this is that the dominant hopping integrals
in plane are between $c$ orbitals, and the largest out of
plane hopping integrals are for $a$ and $b$.
Therefore we have only two coupling constants
$U_\parallel$ and $U_\perp$ describing these 
two physically different interactions.
Our strategy is to adjust these phenomenological parameters
in order to obtain one transition at
the experimentally determined $T_c$.
Thus, beyond fitting $T_c$,
there are no further
adjustable parameters, and one can compare
directly the calculated physical properties of the superconducting
states to those  experimentally observed. Consequently, if
 one obtains a good overall agreement one can say
that one has empirically determined the form of the
pairing interaction in a physically transparent manner.

\begin{figure}[b]
\centerline{\epsfig{file=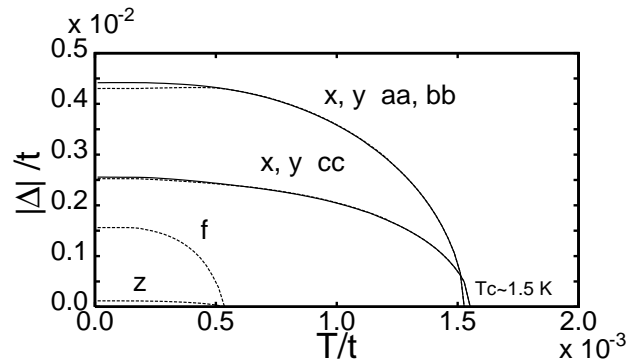,width=4.0cm,angle=-90}}
\vspace{1cm}
\caption{Order parameters, $|\Delta^x_{aa}|$,
$|\Delta^{x}_{cc}|$, $|\Delta^{z}_{aa}|$, 
$|\Delta^{f}_{aa}|$ as functions of temperature,
(dashed lines), and  excluding $z$ and $f$ (full lines).}
\label{fig1}
\end{figure}

Because the pairing interactions 
 $U^{\sigma\sigma'}_{mm'}(ij)$ were assumed to act
only for nearest neighbour sites in or out of plane, 
the pairing potential $ \Delta^{\sigma\sigma'}_{mm'}(ij)$
is also restricted to nearest neighbours.
We further focus on only odd parity (spin triplet)
pairing states for which the vector ${\bf d} \sim (0,0,d^z)$,
i.e. $ \Delta^{\uparrow\downarrow}_{mm'}(ij)=
  \Delta^{\downarrow\uparrow}_{mm'}(ij)$, and
$ \Delta^{\uparrow\uparrow}_{mm'}(ij)=
\Delta^{\downarrow\downarrow}_{mm'}(ij)=0 $.
Therefore in general we have the following non-zero
order parameters  (i) for in plane bonds:
$\Delta_{cc}(\hat{\bf e}_x)$, $\Delta_{cc}(\hat{\bf e}_y)$,
and (ii) for inter-plane bonds: 
$\Delta_{aa}({\bf R}_{ij})$,
$\Delta_{ab}({\bf R}_{ij})$,  $\Delta_{bb}({\bf R}_{ij})$
for ${\bf R}_{ij}=(\pm a/2, \pm a/2, c/2)$.

Taking the lattice Fourier transform of Eq.~\ref{deltas}
the corresponding pairing potentials in k-space
have the general form (suppressing the spin indices for clarity):
\begin{equation}
 \Delta_{cc}({\bf k})  = \Delta_{cc}^x  \sin{k_x} +
 \Delta_{cc}^y  \sin{k_y} \label{eq.delta2}
\end{equation}
for $c$ orbitals and,  
\begin{eqnarray}
&& \Delta_{mm'}({\bf k}) = \Delta^{z}_{mm'} \sin{\frac{k_zc}{2}}
 \cos{\frac{k_x}{2}} \cos{\frac{k_y}{2}} 
\nonumber \\
&&+\Delta^{f}_{mm'}  \sin{\frac{k_x}{2}} \sin{\frac{k_y}{2}}
 \sin{\frac{k_zc}{2}} \\
&&
 +\left(\Delta^{x}_{mm'} \sin{\frac{k_x}{2}} \cos{\frac{k_y}{2}}  
+\Delta^{y}_{mm'} \sin{\frac{k_y}{2}} \cos{\frac{k_y}{2}} \right) 
\cos{\frac{k_zc}{2}} \nonumber
\label{eq.delta2}
 \end{eqnarray}
for $m,m'=a,b$. 
Note that beyond the usual
p-wave symmetry of the $\sin{k_x}$ and $\sin{k_y}$ type 
for the $c$ orbitals, we include all three
additional p-wave symmetries of the $\sin{k/2}$ type which are 
induced by the effective attractive interactions between carriers on the
neighboring out-of-plane Ru orbitals. 
These interactions are also responsible 
for the f-wave symmetry order parameters, $\Delta^{f}_{mm'}$,
transforming as $B_{1u}$ in the notation of \cite{annett}.
This latter is symmetry
distinct from all p-wave order
parameters in a tetragonal crystal,
unlike some other f-wave states which have been 
proposed\cite{graf,won,dahm}. 
The $p_z$ order parameters  $\Delta^{z}_{mm'}$ are of
$A_{2u}$ symmetry. In contrast the pairs 
$\Delta^x_{mm'},\Delta^y_{mm'}$ for $m,m'=a,b$
are of the same $E_u$ symmetry as $\Delta_{cc}^x,\Delta_{cc}^y$.
In general the order parameters in 
each distinct irreducible representations will have 
different transition temperatures.

\begin{figure}[thb]
\centerline{\epsfig{file=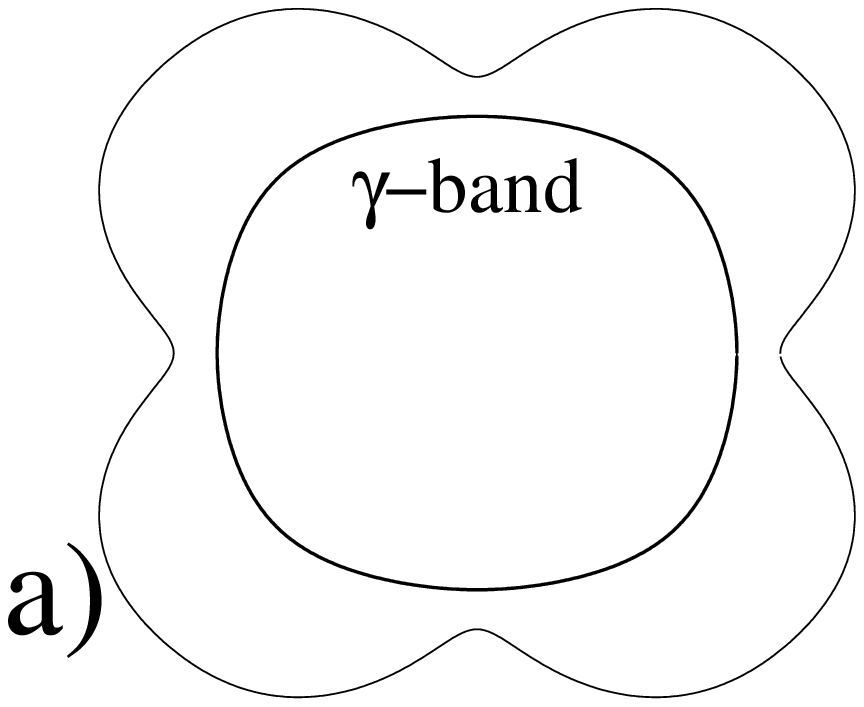,width=3.5cm,angle=0} \hspace{0.5cm}
\epsfig{file=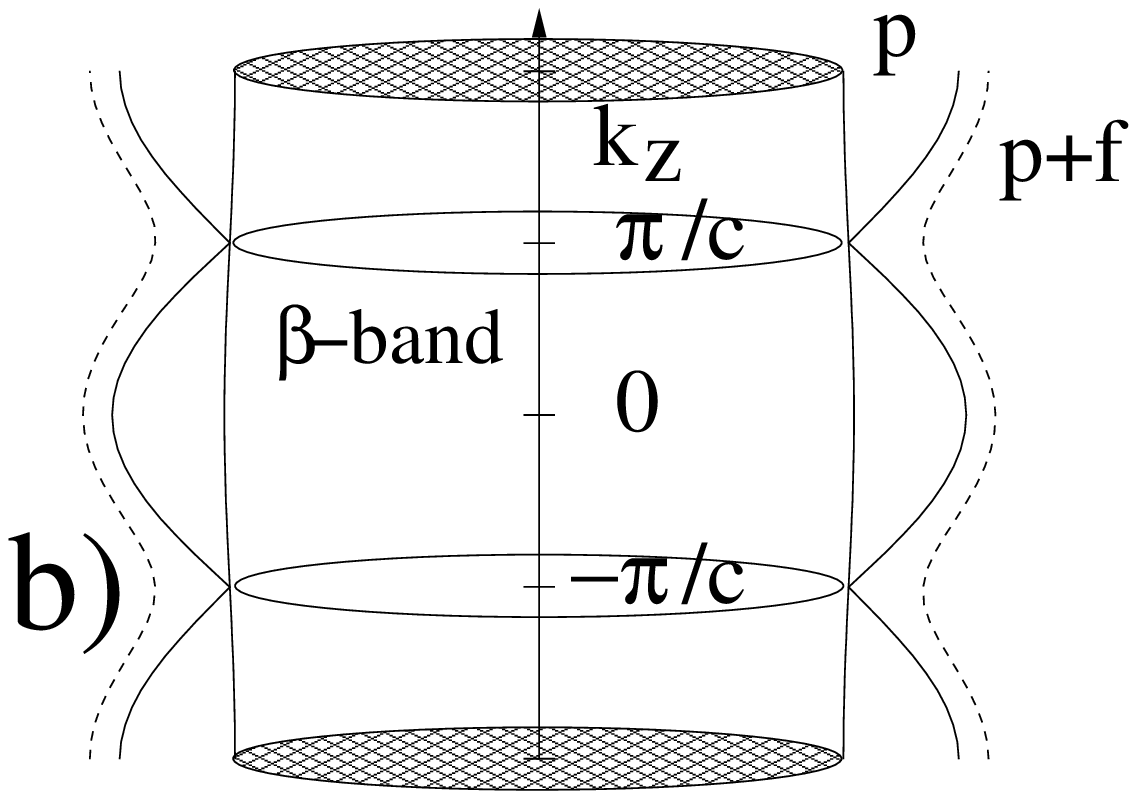,width=4.0cm,angle=0}}
\vspace{1cm}
\caption{Lowest energy eigenvalues, $E^\nu({\bf k})$  
on the Fermi surface; (a)
$\gamma$ sheet 
in the plane $k_z=0$, $E_{\gamma,max}({\bf k}_F)=
0.25{\rm meV}$, $E_{\gamma,min}({\bf k}_F)=0.064{\rm meV}$
(b)  $\beta$ sheet
in the plane $k_x=k_y$, $E_{\beta,max}({\bf k}_F)= 0.32{\rm meV}$.
The $f-$wave order parameter lifts the
p-wave line nodes (dotted lines)}
\label{fig2}
\end{figure}

Fig.~\ref{fig1} shows the calculated order parameters as
a function of temperature. One can see that the $f-$wave and $p_z$
states have a much smaller $T_c$ than the leading
$p_x,p_y$ gap parameters for $aa$, $bb$ and $cc$ orbitals. 
Above the 
$f$-wave $T_c$ the order parameters have the symmetries
$\Delta_{cc}^y=i\Delta_{cc}^x$, 
$\Delta_{bb}^y=i\Delta_{aa}^x$
as expected for a pairing symmetry\cite{agte}
$(k_x+ik_y)\hat{\bf e}_z$
corresponding to the same time reversal broken pairing
state as $^3He-A$.  The off-diagonal components, such as
$\Delta^x_{ab}$ are small but non-zero, 
as are $\Delta_{bb}^x$ and $\Delta_{aa}^y$.
Note that the k-space pairing potentials 
$\Delta_{mm'}({\bf k})$
do not directly correspond to the energy gaps on the Fermi surface sheets,
shown in Fig.~\ref{fig2},
because the tight-binding Hamiltonian is non-diagonal in the orbital 
indices.

\begin{figure}[bh]
\hspace*{1mm}\centerline{\epsfig{file=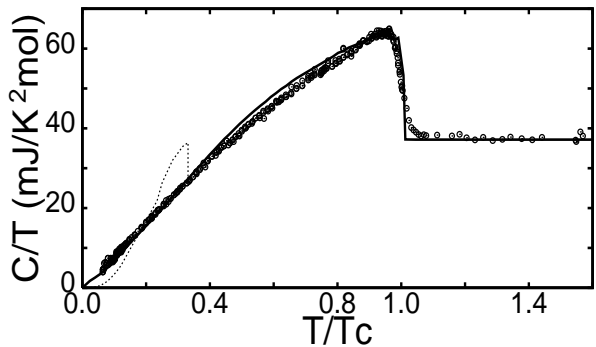,width=4.0cm,angle=-90}}
\vspace{1cm}
\caption{Calculated specific heat,  compared to the experimental data
of NishiZaki {\it et al.}.[14]}
\label{specific}
\end{figure}

Using the above quasi-particle spectra we have calculated the specific
heat, shown in Fig.~\ref{specific}.
Remarkably, although we fitted to $T_c$ only, the
specific heat jump at $T_c$ is  $27$mJ/mol,  in 
essentially exact agreement with the experiment\cite{nishi}.
Thus, contrary to the argument given by Zhitomirsky and Rice\cite{zhit}
it is not necessary to assume a gap opening up on the active
$\gamma$ sheet only in order to
obtain the correct discontinuity at $T_c$. 
Furthermore the calculated specific heat also follows
the experiment closely at lower temperatures, except
for the second transition at about $0.3 T_c$.  
Clearly from Fig.~\ref{fig1} this is due to the f-wave
and $p_z$ pairing, which are symmetry decoupled from the $p_x,p_y$ states,
and  is not consistent with the
experiments\cite{nishi}.   However, if we suppress the 
$f-$wave and $p_z$ pairing then the specific heat
agrees perfectly down to the lowest temperatures.
We believe that this is physically reasonable, because
the f-wave states will be more strongly influenced
by impurities than the p-wave states, due to a $(2l+1)$
prefactor in the impurity pair breaking parameter\cite{impurity}.
As can be seen in Fig.~\ref{fig1} the $x,y$\  $aa$ and $cc$
order paramameters are only slightly affected by the presence of the
f-wave gap.

The low temperature limit of the specific heat is power
law, because our gap parameters have line nodes.
These are horizontal circles around the cylindrical
$\alpha$ and $\beta$ Fermi surface sheets, as illustrated
in Fig.~\ref{fig2},
while the $\gamma$ sheet is node-less. As can be seen, when the f-wave
order parameter also becomes finite, the line nodes disappear.
The fact that the slope of specific heat at low temperatures
(without f) agrees quantitatively with the experiments
suggests that these horizontal nodes are only present on
$\alpha$, $\beta$, as in our model.

\begin{figure}[tb]
\hspace*{1mm}\centerline{\epsfig{file=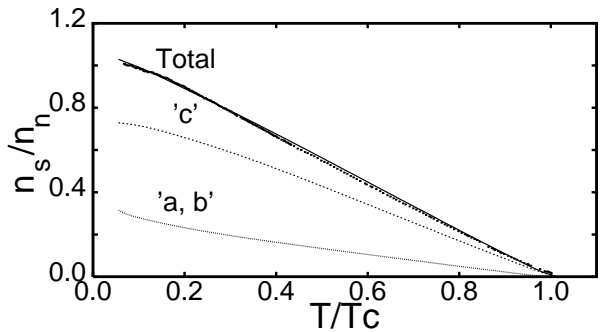,width=4.0cm,angle=-90}}
\vspace{0.5cm}
\caption{Superfluid density as a function of T (solid line),
and experimental points from Sample 1 of Bonalde {\it et al.}[15].
The relative contributions of $c$ and $a,b$
order parameters are indicated (dashed lines).}
\label{fig4}
\end{figure}

A further independent test of our model is the calculation of the
superfluid density\cite{agte}, shown in Fig.~\ref{fig4}. Again there
is excellent agreement over the whole temperature range between $T_c$
and zero. Some physical insight into the different contributions
to the superfluid density can be found by setting
 $\Delta^{x,y}_{mm'}$ to zero for $m=m'=c$ or $mm'=a,b$
and repeating the calculation.
The $c$ orbital only contribution gives about
$70\%$ of the zero temperature superfluid density,
with the remaining $30\%$ derives from the $a,b$ orbital 
order parameters. One can see that the finite slope at zero temperature
derives only from  $a,b$ contributions, consistent with the
the line nodes on the $\alpha,\beta$ Fermi surfaces. 
The only disagreement is that the absolute magnitude of 
our calculated zero temperature x-y plane penetration depth
is only $450\AA$ compared to $1900\AA$ determined
 experimentally\cite{riseman,luke2}.  This discrepancy
may be due to impurities (the experimental samples
had $T_c \sim 1.1-1.3K $ ) or to non-local electrodynamic
effects associated with the line nodes\cite{leggett2}.
We have also calculated the temperature dependent zero-field
thermal conductivity, which is also in good qualitative agreement
with the experiments of Izawa {\it et al.}\cite{izawa}.

In summary we would like to emphasize two points. Firstly, we have proposed an
alternative to the the `intra-band proximity effect' model of 
Zhitomirsky and Rice\cite{zhit} for describing 
horizontal line nodes on the $\alpha ,\beta $ sheets
of the Fermi Surface in superconducting Sr$_{2}$RuO$_{4}$. 
Our bond model
differs from theirs in the way the interlayer coupling is implemented. Our
description is a real space, two point tight-binding\ interaction such as
naturally arises in any multi-band, extended, negative $U$ Hubbard model,
Eq.~\ref{hubbard}. To
be quite clear about this matter we recall that a generic pair-wise
interaction like $U({\bf r},{\bf r}')$, when expressed in the language of 
a tight-binding model Hamiltonian will, in general, give rise to 
four point
interaction parameters $U_{ij,kl}$. The original Hubbard Hamiltonian makes
use of the one point parameters $U_{i}^{(1)}=U_{ii,ii}$ whilst the extended
Hubbard model  is based on two point parameters $U_{i,j}^{(2)}=U_{ij,ij}$.
Evidently our `bond' model is a negative U-version of the
latter\cite{micnas}.  On the other hand
the `proximity effect model'\cite{zhit} corresponds to 
using $U_{ij,i}^{(3)}$.  The
physics of this is often referred to as assisted hopping\cite{zawadowski}.
If one assumes, as
is normally the case in an isotropic substance, that $
U^{(1)}>U^{(2)}>U^{(3)}>U^{(4)}$ then the `bonds' represent stronger
coupling than  assisted hopping  and should be the preferred coupling
mechanism. However, for the tetragonal arrangement of Ru atoms in 
Sr$_{2}$RuO$_4$ this is no more than a suggestion at present. 
Thus the relative merits of the two
models will eventually be settled by appeal to experiments.

Secondly, we wish to stress that in our `bond' approach to the problem the
parameters which describe the normal state are determined by fitting to the
very accurately known Fermi Surface and the measured $T_{c}$ determines
both  coupling constants $U_\parallel=40{\rm meV}$,
$U_\perp=48{\rm meV}$. 
Thus, the calculated
specific heat, Fig.~\ref{specific}, and superfluid density,
Fig.~\ref{fig4}, are parameter free
quantitative predictions of the theory. Consequently,  their  good agreement
with experiments can be construed as strong support for the
 physical picture represented by our model. 
Interestingly a significant feature of this is that
the amplitude of the gap function,and correspondingly the superfluid
density, on the $\alpha ,\beta$ sheets of the Fermi Surface are comparable
to that on the $\gamma$ sheet. This is unlike the case suggested by the 
`inter-band proximity effect' picture\cite{zhit}
where  $\gamma $ is the active
band and $\alpha,\beta $ play a passive role. Hopefully, experiments will
soon clarify which of the two situations prevails.
As for the physical mechanism of pairing, the fact that 
$U_\perp \approx U_\parallel$ implies that the pairing interaction is 
fairly isotropic in spite of the layered structure.


This work has been partially supported by KBN grant No. 2P03B 106 18 and the 
Royal Society Joint Project. We are grateful to Prof. Y. Maeno and 
Prof D. van Harlingen for providing us with 
the experimental data used in Figs. 3 and 4.

\end{document}